\documentclass{pasj00}

\begin{document}
\SetRunningHead{Kitayama et al.}{Exploring Cluster Physics  with 
High-Resolution SZE Images and X-Ray Data}

\Received{2003/10/01}
\Accepted{2003/11/25}
\Published{2004/2/25}

\title{Exploring Cluster Physics with High-Resolution Sunyaev--Zel'dovich
Effect Images and X-Ray Data: ~The Case of the Most X-Ray-Luminous 
Galaxy Cluster RX~J1347--1145}

\author{
Tetsu \textsc{Kitayama},\altaffilmark{1}
Eiichiro \textsc{Komatsu},\altaffilmark{2,3} 
Naomi \textsc{Ota},\altaffilmark{4}
Takeshi \textsc{Kuwabara},\altaffilmark{5}
Yasushi \textsc{Suto},\altaffilmark{5,6}\\
Kohji \textsc{Yoshikawa},\altaffilmark{5} 
Makoto \textsc{Hattori},\altaffilmark{7}
and
Hiroshi \textsc{Matsuo}\altaffilmark{8}
}

\altaffiltext{1}{Department of Physics, Toho University,  
Funabashi, Chiba 274-8510}
\email{kitayama@ph.sci.toho-u.ac.jp}
\altaffiltext{2}{Department of Physics,
Princeton University, Princeton, NJ 08544, USA}
\altaffiltext{3}{Department of Astronomy,
The University of Texas at Austin, Austin, TX 78712, USA}
\altaffiltext{4}{Department of Physics, Tokyo Metropolitan University,
Hachioji, Tokyo 192-0397}
\altaffiltext{5}{Department of Physics, The University of Tokyo,
Tokyo 113-0033}
\altaffiltext{6}{Research Center for the Early University,
The University of Tokyo, Tokyo 113-0033}
\altaffiltext{7}{Astronomical Institute, Tohoku University, Aoba,
Sendai 980-8578}
\altaffiltext{8}{National Astronomical Observatory of Japan, Mitaka, Tokyo
181-8588}

%


\maketitle

\begin{abstract}
Foreseeing the era of high spatial resolution measurements of the
Sunyaev--Zel'dovich effect (SZE) in clusters of galaxies, we present a
prototype analysis of this sort combined with Chandra X-ray data. It is
applied specifically to RX~J1347--1145 at $z=0.451$, the most
X-ray-luminous galaxy cluster known, for which the highest resolution
SZE and X-ray images are currently available. We demonstrate that the
combined analysis yields a unique probe of complex structures in the
intracluster medium, offering determinations of their temperature,
density, and line-of-sight extent.  For a subclump in RX~J1347--1145,
previously discovered in our SZE map, the temperature inferred after
removing the foreground and background components is well in excess of
20 keV, indicating that the cluster has recently undergone a violent
merger.  Excluding the region around this subclump, the SZE signals in
submillimeter to centimeter bands (350, 150, and 21~GHz) are all
consistent with those expected from Chandra X-ray observations.  We
further present a temperature deprojection technique based on the SZE
and X-ray images, without any knowledge of spatially resolved X-ray
spectroscopy. The methodology presented here will be applicable to a
statistical sample of clusters available in the future SZE surveys.
\end{abstract}

\section{Introduction}

In the era of precision cosmology represented by the first-year WMAP
data (Bennett et al. 2003; Spergel et al. 2003) among others, the
scientific goals of studies on galaxy clusters are gradually shifting
from the exploration of cosmology to investigations of cluster physics
itself.  For cosmological uses, clusters of galaxies are often assumed
to be idealized spherical and homogeneous systems. Recent X-ray
observations by Chandra and XMM-Newton, however, have revealed complex
structures in a number of clusters, whose origins are still to be
clarified (e.g. Markevitch et al. 2000, 2002; Peterson et al. 2003;
Fabian et al. 2003).  There has hence been an increasing need for
accurate and reliable measurements of clusters at multi-wavelengths to
further pursue studies on cluster physics.

The Sunyaev--Zel'dovich effect (SZE), the inverse Compton scattering of
the cosmic microwave background by hot electrons, serves as yet another
diagnosis of the intracluster medium (Sunyaev, Zel'dovich 1970, 1972; for
reviews see Rephaeli 1995; Birkinshaw 1999; Carlstrom et al. 2002). The
intensity of the thermal SZE is proportional to the line-of-sight
integral of the electron density times the temperature, $\int n_{\rm e}
T_{\rm e} dl$, whereas that of the X-ray thermal Bremsstrahlung emission
is proportional to $\int n_{\rm e}^2 T_{\rm e}^{1/2} dl$.  The
combination of the two thus yields independent measures of the density
and temperature distributions, including the departure from spherical
symmetry (Zaroubi et al. 1998; Hughes, Birkinshaw 1998; Yoshikawa, Suto
1999; Donahue, et al. 2003).  In addition, it is still not easy to
explore the nature of very hot ($> 10$~keV) gas with X-ray facilities
alone, as the energy band of X-ray spectrometers with sub-arcminute
spatial resolution is currently limited to below 10~keV.  The SZE has an
advantage in resolving such hot gas even in distant clusters, given its
linear dependence on $T_{\rm e}$ and its redshift independence
(e.g. Rephaeli 1995).  This is of particular importance in depicting
violent activities in clusters, such as mergers and associated heating
of the gas.

As of today, the spatial resolution of the majority of the SZE imaging
observations is limited to arcminute scales, considerably lower than
current X-ray observations ($6''$ for XMM-Newton, $0''.5$ for
Chandra). This is largely due to the limited number of radio telescopes
with sub-arcminute resolution in the range of wavelengths relevant to
the SZE measurements.  Controlling the systematics of such large
telescopes to the sensitivity level required for the SZE imaging is also
challenging.  Interferometers are stable with carefully controlled
systematics and have allowed high signal-to-noise ratio (S/N) imaging of
the SZE (e.g., Jones et al. 1993; Carlstrom et al. 1996); the effective
resolution ranges from $15''$ to a few arcmin, with more S/N at the
largest angular scales. A single-dish telescope with an array of
detectors working at millimeter or submillimeter wavelengths is also a
promising tool of the SZE imaging observations. It is highly plausible
that an array with hundreds to thousands of bolometers, such as BOLOCAM
(Glenn et al. 2003) and SCUBA-2 (Holland et al. 2003), will achieve an
angular resolution of $\sim 10''$ while retaining a wide field-of-view
of $\sim 10'$.

Given the recent progress and numerous on-going plans of the SZE
surveys, it is likely that future cluster data, especially at high
redshifts, will be dominated by the SZE images and supplemented by the
X-ray and optical follow-up observations. Therefore, we explore uses of
the SZE data beyond the standard analysis done so far.  An ideal and
unique template for this purpose is provided by RX~J1347--1145 at
redshift $z=0.451$, the most X-ray-luminous cluster of galaxies known
(Schindler et al. 1995, 1997).  This cluster is an exceptional target of
high-resolution SZE observations for two reasons: 1) the central SZE
intensity, characterized by the central $y$-parameter of $y_0 \sim
10^{-3}$, is more than twice as large as any other clusters observed to
date, and 2) its compact size, characterized by an angular core radius
of $\theta_{\rm c} \sim 7''$, enables mapping within a relatively narrow
field-of-view of existing facilities.  A number of measurements have
thus been carried out for this cluster (Komatsu et al. 1999, 2001;
Pointecouteau et al. 1999, 2001; Reese et al. 2002).

We currently have SZE maps of RX~J1347--1145 at three bands: an SZE {\it
increment} map at 350~GHz with a spatial resolution of $15''$ (Komatsu
et al. 1999), and {\it decrement} maps at 150~GHz and 21~GHz with the
resolutions $13''$ and $76''$, respectively (Komatsu et al. 2001).  In
particular, our 150~GHz observation has achieved the highest resolution
of the SZE images so far and revealed a complex substructure with
enhanced decrement signals at $\sim 20''$ off the X-ray center.  The
substructure was not visible in the previous ROSAT X-ray image
(Schindler et al. 1997), indicating that it has a very high temperature
that is far beyond the energy band of the ROSAT/HRI ($<2$~keV).  This
finding was subsequently confirmed by the Chandra observations with an
emission-weighted temperature toward the substructure of
$18.0^{+2.7}_{-2.3}$~keV, much higher than the ambient temperature $12.7
\pm 1$~keV (Allen et al. 2002).  Based on their Keck spectroscopy of 47
member galaxies, Cohen and Kneib (2002) suggested that this cluster is
undergoing a major merger.  Indeed, this is a good demonstration that the
SZE imaging observations provide an independent and powerful diagnosis
of cluster physics.

In this paper, we combine our multi-band, high-resolution SZE images and
the Chandra X-ray data of RX~J1347--1145 to present a proto-type
analysis that will continue to prove useful in the future. More
specifically, we take into account radial temperature profile and
non-sphericity of the intracluster gas to constrain the gas distribution
and the cluster bulk motion.  We also attempt to deproject temperature
profile of the cluster on the basis of the SZE and X-ray surface
brightness, without a knowledge of spatially resolved X-ray
spectroscopy.  Finally, we explore the physical nature of the
high-temperature substructure and discuss its implications on the merger
history and dynamical activities of the cluster.

Throughout the paper, we assume a standard set of cosmological
parameters: $\Omega_m=0.27$, $\Omega_\Lambda=0.73$, and $h=0.71$
(Spergel et al. 2003). In this cosmology, an angular size of 1$''$
corresponds to a physical size of 5.74 kpc at the cluster redshift
$z=0.451$. Unless stated otherwise, errors are given in 1-$\sigma$.

\begin{table*}
\caption{Summary of the SZE observations and the data used in the
 analysis. }
\label{tab:obs}
\begin{center}
\begin{tabular}{cccc}
\hline \\[-9pt] 
\hline \\[-6pt] 
Frequency & 350 GHz& 150 GHz& 21 GHz\\[4pt]\hline \\[-6pt]
Facility & JCMT/SCUBA & Nobeyama/NOBA& Nobeyama/HEMT \\ 
Beam-size (FWHM)  &$15''$ & $13''$ & $76''$ \\  
Field-of-view & $2'.8$ diameter  & $1'.9 \times 1'.9$ & $6'.0 \times 6'.0$\\
Average 1-$\sigma$ noise 
& 5.3 mJy~beam$^{-1}$& 1.6 mJy~beam$^{-1}$& 0.9 mJy~beam$^{-1}$\\
Radial bins [arcsec] & 10--20, 20--30, 30--40, & 10--20, 20--30, 30--40,
 & 60--110, 110--140, \\
& 40--50, 50--65, 65--85&  40--50, 50--60, 60--75& 140--170, 170--240  
\\[4pt] \hline 
\end{tabular} 
\end{center}
\end{table*}

\section{Multi-Wavelength Data of RX~J1347--1145}

We describe below the observations and the data used in the present
paper.  The summary of the SZE data is presented in table \ref{tab:obs}.

\subsection{Mapping Observations with SCUBA at 350~GHz}

Our submillimeter mapping observations at 350~GHz were performed with
the Sub-millimeter Common User Bolometer Array (SCUBA) (Holland et
al. 1999), attached on the Nasmyth platform of the James-Clark-Maxwell
Telescope (JCMT) in Hawaii.  There are 37 and 91 bolometers at 350 and
650~GHz, respectively.  The observations were carried out in 1998 and
1999. The first year data are published in Komatsu et al. (1999). The
radial profile of the first year data clearly indicated an extended
positive signal consistent with the SZE increment feature, and the
fitted amplitude of the SZE matches the prediction of the X-ray data.
While the azimuthally averaged radial profile had a reasonable S/N, the
noise per each pixel of the map was too high to study the detailed
morphology of the SZE, owing to poor weather. A typical zenith optical
depth during the first year was $\tau_{350}=0.46-0.60$, which
corresponds to the ``wet'' condition at JCMT. The 1-$\sigma$ noise level
averaged over the first year map was 8.0~mJy~beam$^{-1}$.

\begin{figure*}
  \begin{center}
    \FigureFile(146mm,146mm){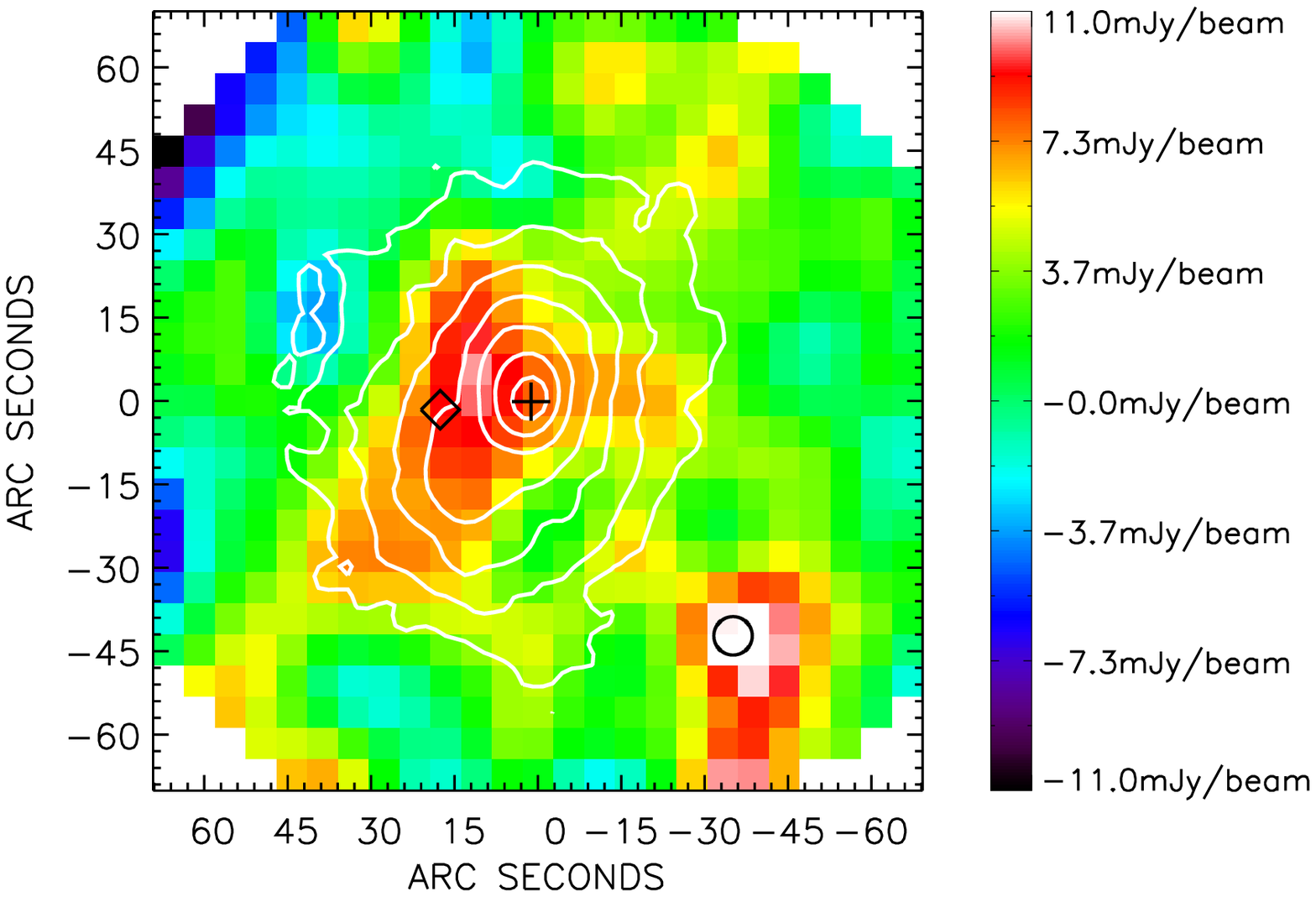}
    \FigureFile(146mm,146mm){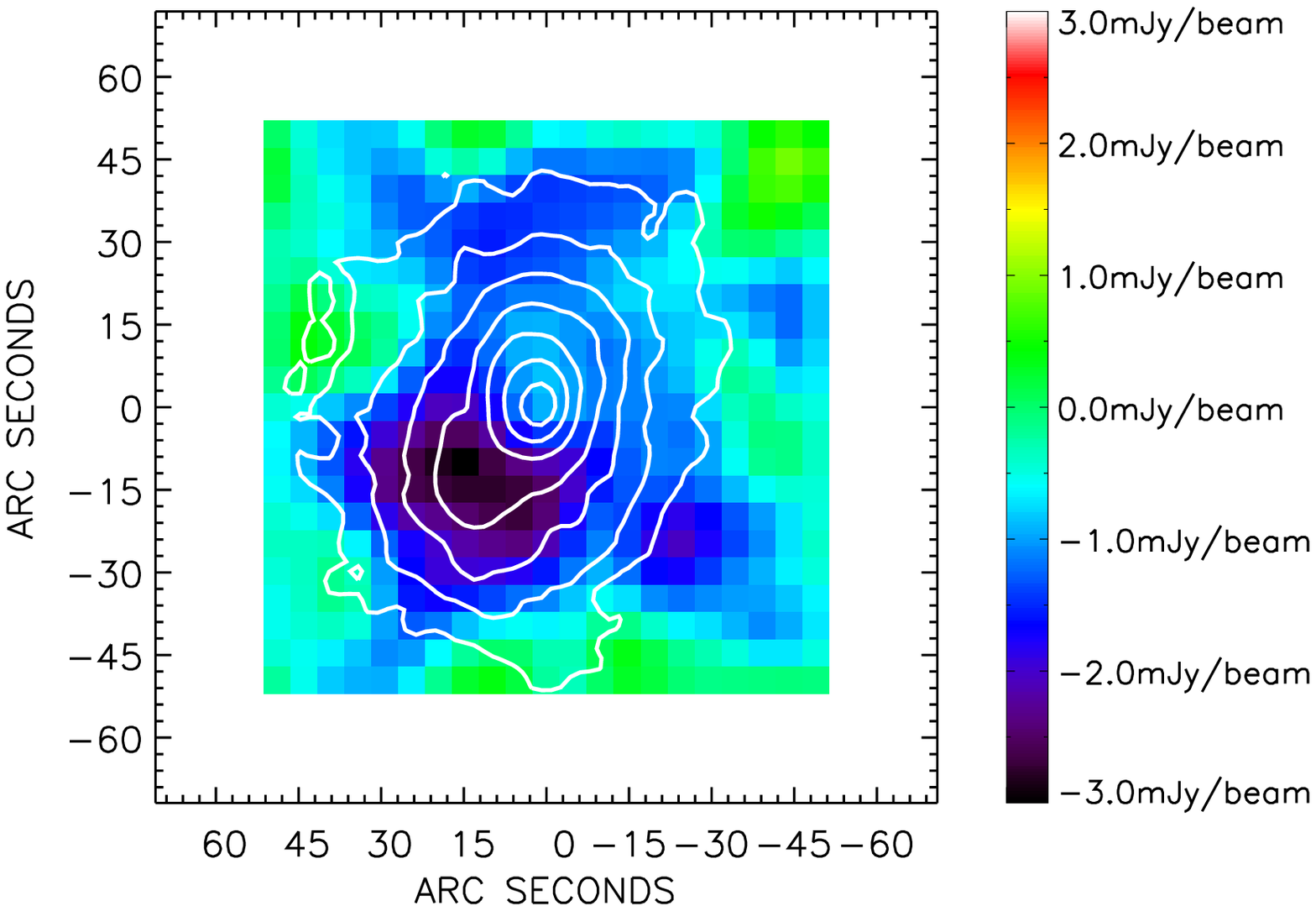}
  \end{center}
  \caption{SZE increment and decrement images of RX~J1347--1145 at
350~GHz (top) and 150~GHz (bottom), respectively.  Both images are
centered at $13^{\rm h} 47^{\rm m}30^{\rm s}.54 -11^{\circ} 45' 09''.4$
(J2000) and smoothed with a 15$''$ Gaussian filter. The rms levels
measured on these smoothed images are 3.0 mJy~beam$^{-1}$ and 0.7
mJy~beam$^{-1}$ at 350~GHz and 150~GHz, respectively.  The positions of
three sources in the field-of-view are marked by a cross (central cD), a
diamond (secondary cD), and a circle (newly-discovered submillimeter
source).  Their contributions are not corrected in the images, but the
pixels containing the sources are excised in the analysis.  Contours
indicate the 0.5--7.0 keV X-ray surface brightness taken by Chandra,
corresponding to 64, 32, 16, 8, 4, 2, and 1\% of the peak value.}
\label{fig:szmaps}
\end{figure*}

In order to improve on the S/N of the map, we performed additional
observations on 1999 July 3 and 4. The weather condition of the second
year was much better than that of the first year mentioned above. A
typical zenith optical depth was $\tau_{350}=0.21$ on July 3 and
$\tau_{350}=0.34$ on July 4. The 1-$\sigma$ noise level averaged over
the second year map is 6.4~mJy~beam$^{-1}$, and that of the coadded map
(shown in figure \ref{fig:szmaps}) is 5.3~mJy~beam$^{-1}$.  We used the
planet Uranus for the primary flux calibration and beam measurements. We
also observed the planet Mars as a secondary flux calibrator, and
checked the stability of the gain of the telescope. There was no
significant gain variation during the observations; thus, the
calibration error is dominated by the uncertainty in the flux of Uranus,
which is less than 15\%. The beam pattern was fitted by an elliptical
Gaussian profile with axis sizes $15''.5 \times 14''.3$. For simplicity,
however, we approximate the beam as a symmetric Gaussian whose size is
the geometric mean of the elliptical fit, $14''.9$.  Since the beam-size
of the first year map was $15''.2$, we use $15''$ as the effective FWHM
beam-size of the coadded map. This approximation of the beam does not
cause significant systematic errors compared to the instrumental noise
level in the coadded map.

In both years, sky chopping was done in the azimuth direction with a
length of $120''$. The observations were made in the ``jiggle'' mode,
where each bolometer samples 64 independent points on the sky separated
by $3''.09$. We have reduced the time-ordered data with the reduction
tools in the {\sf SURF} package (Jenness, Lightfoot 1998), following the
standard data analysis pipeline. After deswiching, flat-fielding,
extinction corrections, and the sky noise correction with {\sf REMSKY}
(Jenness et al. 1998), we have removed the spikes above 4-$\sigma$ in
the time-ordered data using {\sf DESPIKE}. The map-making routine, {\sf
REBIN}, allocates the reduced time-ordered data onto a rectangular grid
with a Gaussian low-pass filtering. We choose $6''$ as a nominal pixel
size in the final map, because it is close to the Nyquist sampling at
350~GHz; thus, each pixel in the map is treated as being
independent. The non-uniformity of noise in the map is taken into
account explicitly.

Measurements of the SZE are in general difficult with SCUBA, except for
very compact clusters, such as RX~J1347--1145. Since the SZE signal is
typically extended across the field-of-view ($\sim 150''$ in diameter),
a fraction of it will be lost by sky chopping.  In the standard data
analysis pipeline, {\sf REMSKY} subtracts the average of some (if not
all) selected bolometers from each bolometer upon each integration,
further reducing the atmospheric noise as well as the SZE signal. Zemcov
et al. (2003) proposed a way to partially resolve this problem; instead
of subtracting the average flux, they subtract the signal correlated
with that at 650~GHz upon each integration.  Their method removes less
SZE and is suitable for extended clusters, because the contribution of
the SZE to the total flux is much smaller at 650~GHz than 350~GHz. It
is, however, reported to increase noise by 15\% compared to {\sf
REMSKY}. Considering both the advantages and disadvantages of the two
methods, we adopted the standard pipeline mentioned above to yield a map
with a better S/N for this particular cluster.

\subsection{Mapping Observations with NOBA at 150~GHz}

The Nobeyama Bolometer Array (NOBA, Kuno et al. 1993), mounted on the
Nobeyama 45-m telescope in Japan, consists of 7 bolometers.  Since the
data we present here are exactly the same as those published in Komatsu
et al. (2001), we outline our millimeter observations at 150~GHz only
briefly, with particular emphasis on differences between the data
reduction procedures of NOBA and SCUBA.

A notable difference between NOBA and SCUBA is that NOBA does not
require chopping. The bolometers are read-out through six differential
circuits between the central bolometer and the other six surrounding
ones; the outputs are differential signals between the central bolometer
and each of the six surrounding bolometers. The atmospheric noise is
thus subtracted electronically rather than mechanically.  In the
map-making algorithm, the six differential outputs are used to
reconstruct signals at the central bolometer. By rotating the array
clockwise by $19^\circ .1$, a scan path yields seven equally-spaced
lines with the separation of $5''.3$, which determines the size of a
pixel.  We have performed two orthogonal raster scans ($X$ and $Y$
scans), each being $111''.3$ wide and consisting of three parallel
scans.

The baseline level varies from one scan to the other due to the
atmospheric noise (the scanning effect). The scanning effect can be
reduced by combining the two orthogonal raster scans. We use the
so-called PLAIT method (Emerson, Gr\"ave 1988), which reduces the
scanning effect in Fourier space by suppressing the low-frequency modes
that are aligned with the scan direction. The final map has the FWHM
beam-size of $13''$, the 1-$\sigma$ noise level of 1.6~mJy~${\rm
beam}^{-1}$, and the field-of-view of $111''.3 \times 111''.3$.

\subsection{Mapping Observations at 21~GHz}

In addition to the high-resolution observations mentioned above, we also
performed mapping observations at 21~GHz using the dual circular
polarization HEMT amplifier mounted on the Nobeyama 45-m telescope.
Details of the observations and the data are presented in Komatsu et
al. (2001).  The map has the FWHM beam-size of $76''$, 1-$\sigma$ noise
level of 0.9~mJy~${\rm beam}^{-1}$, and field-of-view of $6'.0 \times
6'.0$.  The capability of this telescope for SZE observations has also
been demonstrated for several other clusters at 21, 36, and 43~GHz
(Tsuboi et al. 1998, 2002; Pointecouteau et al. 2002).

\subsection{Radio Sources} 
\label{sec:sources}

Radio sources can be serious contamination against measurements of the
SZE (e.g., Cooray et al. 1998). In our analysis, we excise pixels that
are potentially contaminated by radio sources.  There are in total three
sources in our observing field; one known radio source, one
newly-discovered submillimeter source, and one candidate source. We
briefly describe their properties below.

At the center of our target cluster, there is a known radio source whose
flux has been measured at several frequencies (Komatsu et al. 1999, 
2001; Pointecouteau et al. 2001).  This source is the central cD galaxy,
and the optical spectroscopy implies that it hosts an AGN with a very
strong OII emission line (Schindler et al. 1995; Cohen, Kneib 2002).
The spectrum in radio bands continues to decline with frequency,
implying synchrotron emission from the AGN. The flux measured by VLA at
22.46~GHz is $11.55 \pm 0.17$ mJy (Komatsu et al. 2001). A power-law
extrapolation of the lower frequency measurements implies $4.3 \pm 0.3$
mJy and $2.7 \pm 0.4$ mJy at 150~GHz and 350~GHz, respectively.

We have also found a bright submillimeter source in our SCUBA image
(figure \ref{fig:szmaps}), located at $\sim 60''$ southwest from the
cluster center.  The flux at 350~GHz is $15.4 \pm 5.3$ mJy. The
existence of the source has been confirmed by our VLA follow-up
observations at 8.46~GHz.  This newly-discovered submillimeter source
merits further investigation, and the results will be reported elsewhere
(I.~Tanaka et al. in preparation).

The secondary cD galaxy located at $18''$ east of the cluster center
(Schindler et al. 1995) may also contaminate our 350~GHz image.  The
optical spectroscopy implies that this source is a red elliptical galaxy
with no OII emission line, and has a comparable $R$-band luminosity to
the central cD (Schindler et al. 1995; Cohen, Kneib 2002). No radio
counterpart has been detected between 1.4~GHz and 150~GHz, suggesting
that this source does not host an AGN.  In our 350~GHz image, an
enhanced emission is indicated around the position of the secondary cD.
This emission may originate from dust in the source and it needs to be
confirmed by infrared observations. Though this source is not
responsible for all of the extended emission seen in the map, it may
distort the image.  To be conservative, we exclude the pixels around
this galaxy in our analysis.

\subsection{SZE Increment and Decrement Images}

Figure \ref{fig:szmaps} shows high-resolution SZE increment and
decrement images we obtained at 350~GHz and 150~GHz, respectively. Both
images are centered at $13^{\rm h} 47^{\rm m}30^{\rm s}.54 -11^{\circ}
45' 09''.4$ (J2000).  For display purposes, they are smoothed by a
Gaussian filter with an FWHM of $15''$, while the intensities are still
given per beam-sizes listed in table \ref{tab:obs}. \footnote{The
definition of the intensity after smoothing is different from that in
figure 1 of Komatsu et al. (2001), which was recalibrated to the
intensity per beam-size of $20''.6$ (an effective beam-size of the
smoothed image at 150~GHz). The significance levels are identical in
both cases. All scientific analyses are done on raw maps before
smoothing and not affected by the above difference.}  The rms levels
measured on these {\it smoothed} images are 3.0 mJy~beam$^{-1}$ and 0.7
mJy~beam$^{-1}$ at 350~GHz and 150~GHz, respectively. The contributions
of the three sources described in subsection \ref{sec:sources} are
explicitly shown. The zero level is defined by the average intensity at
the map edge. Overlaid are the contours of 0.5--7.0 keV X-ray surface
brightness taken by Chandra.

As first discovered by Komatsu et al. (2001), the SZE decrement at
150~GHz shows a clear enhancement at $\sim 20''$ southeast from the
X-ray center of this cluster. The significance of the southeast peak in
the 150~GHz image is at the 4.2-$\sigma$ level, strongly indicating that
there is a substructure in this region. The position of the peak is in
good agreement with that of the enhanced X-ray emission found by Chandra
(Allen et al. 2002).  The increment image at 350~GHz also shows a
similar extended feature in the southeast region, while a higher noise
level and possible contamination from the secondary cD make it harder to
identify the excess emission by eye.

\begin{figure*}
  \begin{center}
    \FigureFile(82mm,82mm){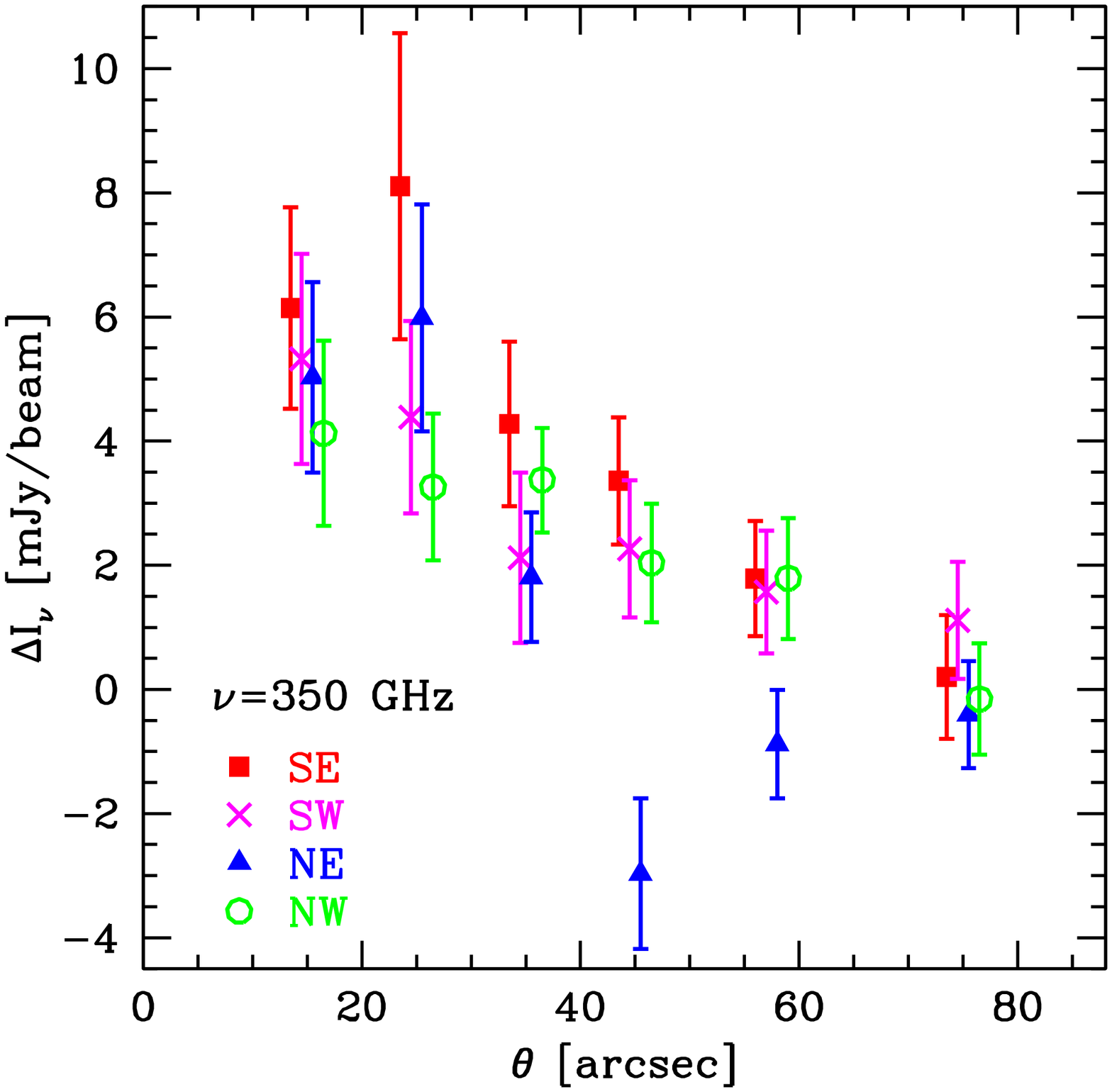}
    \FigureFile(82mm,82mm){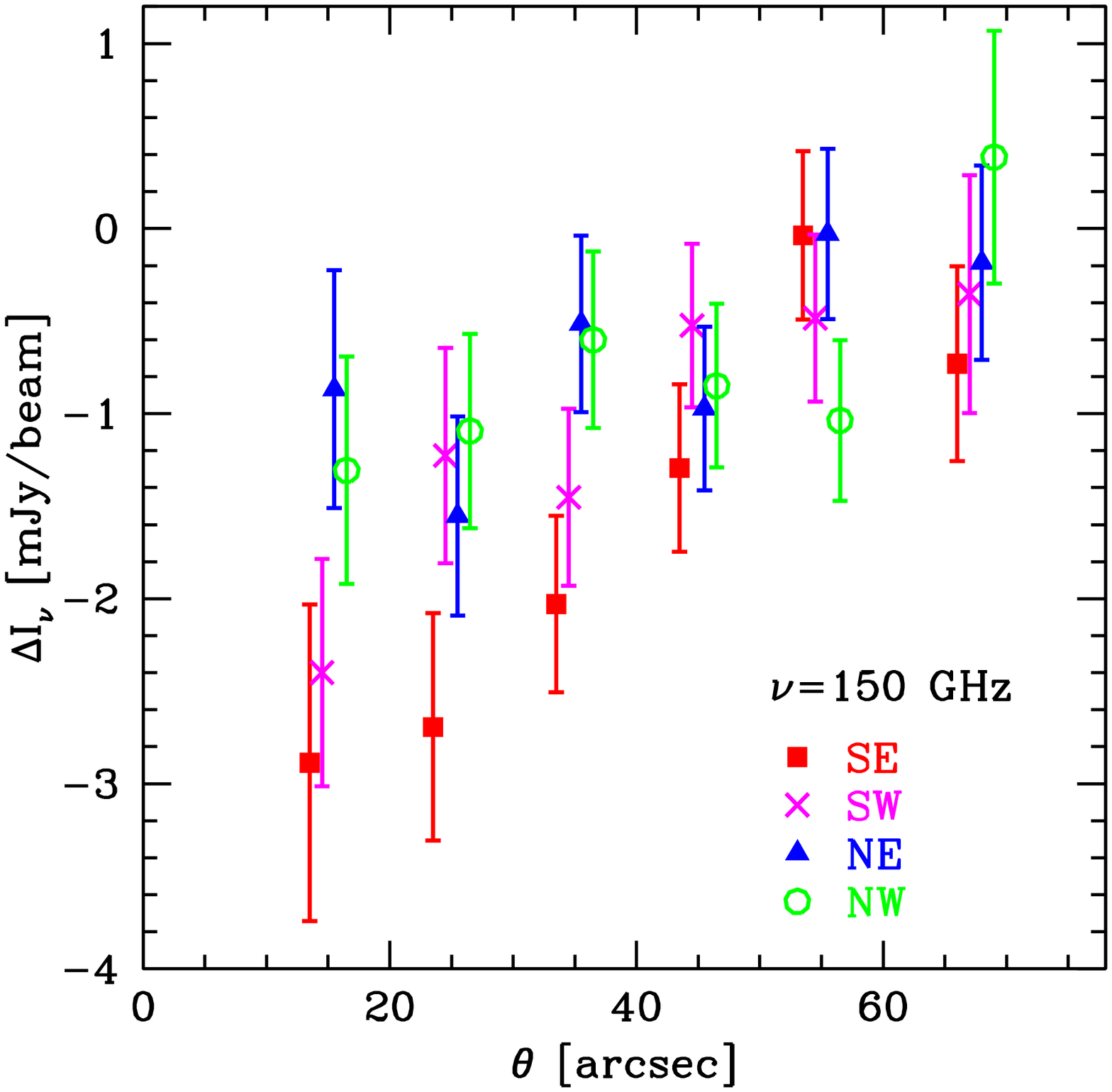}
  \end{center}
  \caption{Radial surface brightness profiles at 350~GHz (left) and
150~GHz (right) in the four quadrants; SE (filled squares), SW
(crosses), NE (filled triangles), and NW (open circles).  The pixels
within a radius of $10''$ from three point sources are excluded. The
zero level is fixed by the average intensity in the outermost bins other
than the SE quadrant.  For clarity, the radial positions are slightly
shifted and horizontal error bars are omitted.}  \label{fig:quad}
\end{figure*}

Although the S/N of the present data is not sufficient for a detailed
pixel-to-pixel confrontation, a quantitative comparison is still
possible by means of azimuthally averaged surface brightness profiles.
We divide the 350~GHz and 150~GHz maps into four quadrants as in figure
2 of Komatsu et al. (2001): southeast (SE), southwest (SW), northeast
(NE) and northwest (NW). Figure \ref{fig:quad} displays radial profiles
in these quadrants after excluding the pixels within a radius of 10$''$
from the three radio sources described in subsection
\ref{sec:sources}. The radial bins are as listed in table \ref{tab:obs}.
To eliminate ambiguities in their absolute values (e.g., by the sky
noise), the SZE signals relative to the map edge are used in our
analysis. More specifically, the zero level of the radial profile data
is fixed by the average intensity in the outermost bins in the SW, NE, and
NW quadrants. The error associated with this zero-level correction is
included in the error assigned to each radial bin.

The detected signals in all the four quadrants in figure \ref{fig:quad}
show extended increment and decrement features characteristic of the SZE
at 350~GHz and 150~GHz, respectively. The signals in the SE quadrant
between radii 10$''$ and 50$''$ are systematically higher than those in
the other directions in both bands. Those in the other directions are on
the whole consistent with one another. Only the $40''-50''$ bin in the
NE quadrant at 350~GHz deviates from the data in the other directions at
the 2.5-$\sigma$ level.  This may partly be due to a reference
(negative) signal of the 15~mJy source in the SW region as a result of
the $120''$ chopping, yet it should diminish below the current noise
level by a sky rotation. We have checked that our results are
insensitive to the presence of this particular bin.

\begin{figure*}
  \begin{center}
    \FigureFile(82mm,82mm){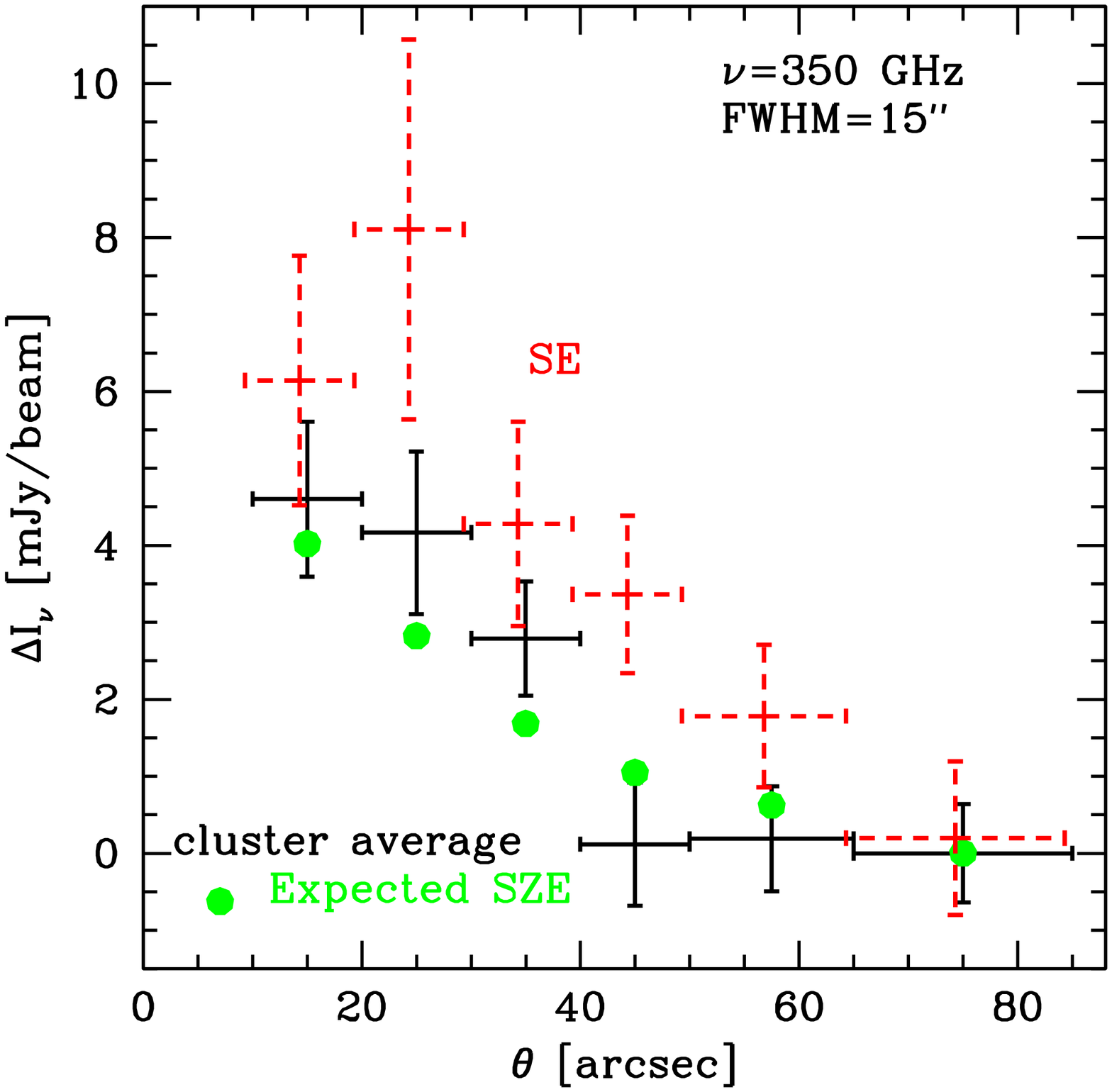}
    \FigureFile(82mm,82mm){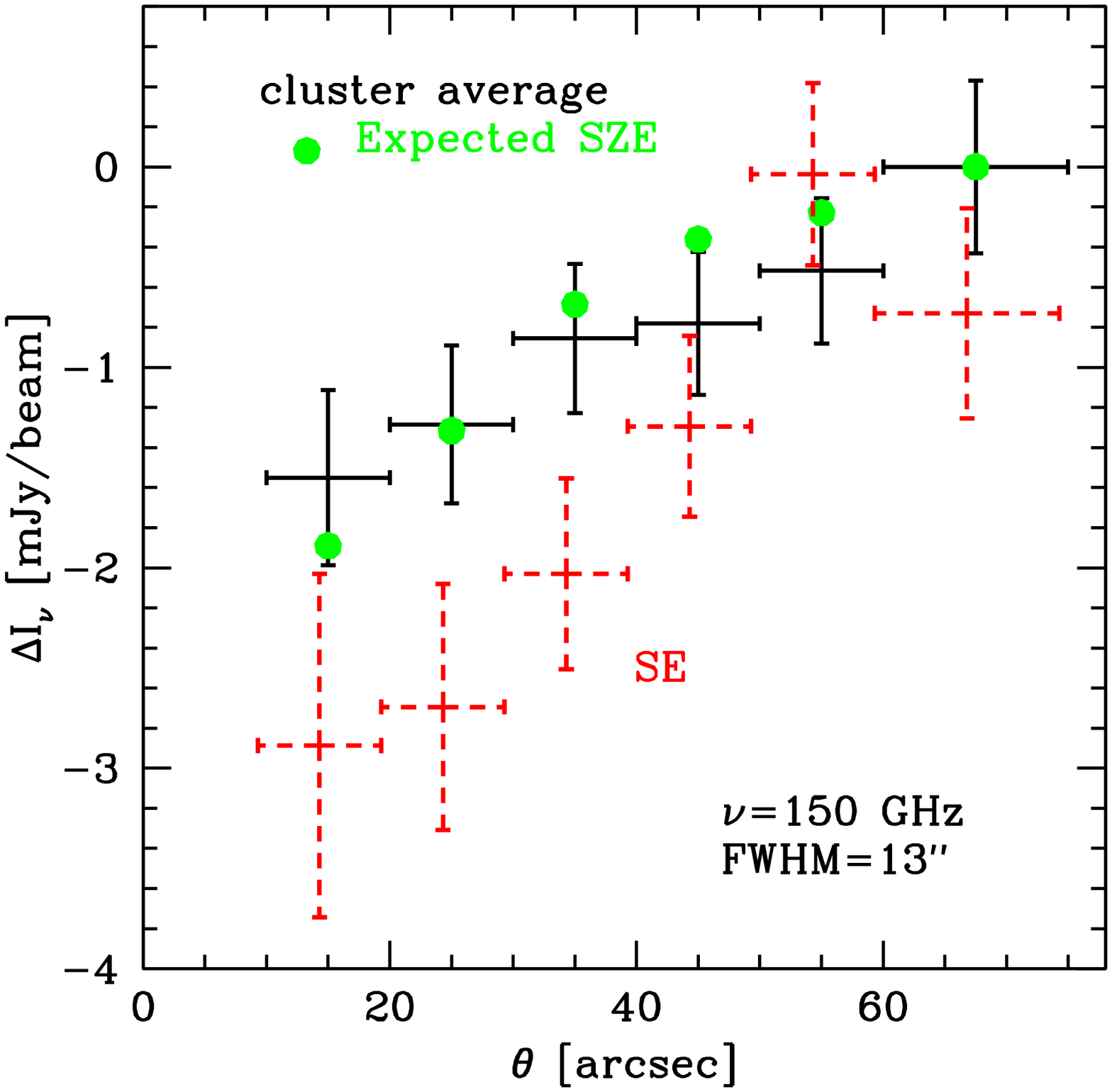}
  \end{center}
  \caption{Same as figure \ref{fig:quad} except for showing the cluster
average component (solid error bars) and of the SE quadrant (dashed
error bars).  Filled circles indicate the SZE signals expected for the
cluster average component from Chandra X-ray imaging and spectroscopy in
the case of spherical gas distribution and zero peculiar velocity (see
section \ref{sec:analy} for details).}  \label{fig:prof}
\end{figure*}

As noted by Allen et al. (2002), the X-ray surface brightness profile is
also regular except for the SE quadrant. In the rest of this paper, we
combine the NW, NE, and SW quadrants and refer to them as the ``cluster
average component''.  Figure \ref{fig:prof} illustrates that the radial
surface brightness profiles of the cluster average component at 350~GHz
and 150~GHz are both consistent with those expected from Chandra X-ray
imaging and spectroscopy of the same region (see section \ref{sec:analy}
for details).  The difference between the signals in the SE quadrant and
the cluster average peaks at $\theta \sim 25''$.

\begin{figure}
  \begin{center}
    \FigureFile(82mm,82mm){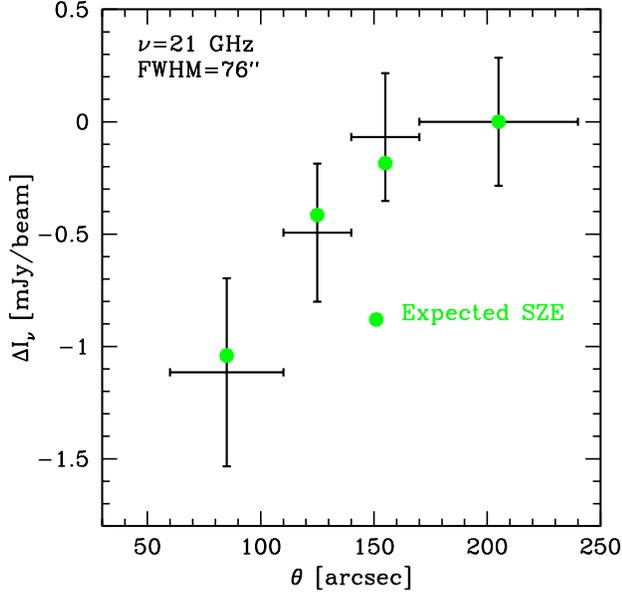}
  \end{center}
  \caption{Radial surface brightness profile at 21~GHz after subtraction
 of the central radio source. The pixels within a radius of $85''$ from
 the center of the excess SZE in the SE region are also excluded.  The
 zero level is determined by the intensity in the outermost bin.  Filled
 circles indicate the SZE signals expected from Chandra X-ray imaging
 and spectroscopy in the case of spherical gas distribution and zero
 peculiar velocity (see section \ref{sec:analy} for details).}
 \label{fig:prof_21}
\end{figure}

Figure \ref{fig:prof_21} shows the radial surface brightness profile of
the 21~GHz data.  The zero level is determined by the intensity in the
outermost bin.  The central source contribution was subtracted using a
direct measurement ($11.55 \pm 0.17$ mJy) at 22.46~GHz.  Because of the
wide beam-size, we do not treat the SE quadrant separately for the
21~GHz data. We instead have excluded the pixels within a radius of
$85''$ from the center of the SE excess indicated in the 150~GHz
image. Consequently the 21~GHz data within $60''$ from the cluster
center are not used in the analysis.  The extended decrement in figure
\ref{fig:prof_21} is also consistent with that expected from the Chandra
observation for the cluster average component.

In summary, we use the radial profile data at 350~GHz, 150~GHz, and
21~GHz presented in figures \ref{fig:prof} and \ref{fig:prof_21}.  At
350~GHz and 150~GHz, we exclude the pixels within a radius of $10''$
from the three radio sources described in subsection \ref{sec:sources}
and treat the SE quadrant separately from the other directions. At
21~GHz, we excise the pixels within a radius of $85''$ from the center
of the SE excess.  In all cases, we measure the intensity differences
relative to the outermost bin.

\subsection{X-Ray Imaging and Spectroscopy with Chandra}
\label{sec:X-ray}

Allen et al. (2002) present their Chandra ACIS observations of
RX~J1347--1145. They found an excess X-ray emission at a position
consistent with the SZE enhancement and unambiguously confirmed the
presence of the substructure reported in Komatsu et al. (2001). Using
the 0.5--7.0 keV band spectra, they determined the emission-weighted
temperature of the SE quadrant between radii $8''.8$ and $28''.6$ as
being $kT_{\rm e} = 18.0^{+ 2.7}_{-2.3}$ keV. This is significantly
higher than the temperature $kT_{\rm e} = 12.7\pm 1.0$ keV of the same
radial bin in other directions. When the SE quadrant is removed, the
X-ray emission appears to be smooth; the mean temperature is $kT_{\rm e}
= 12.0^{+0.62}_{-0.59}$~keV and the mean metallicity is $Z=0.41\pm
0.07$~Z$_\odot$ in the $0''-108''.2$ region. The X-ray spectra indicate
clear evidence of a temperature gradient (figures 4 and 5 of Allen et
al. 2002) and marginal evidence for a metallicity gradient.

\begin{figure}
  \begin{center}
    \FigureFile(82mm,82mm){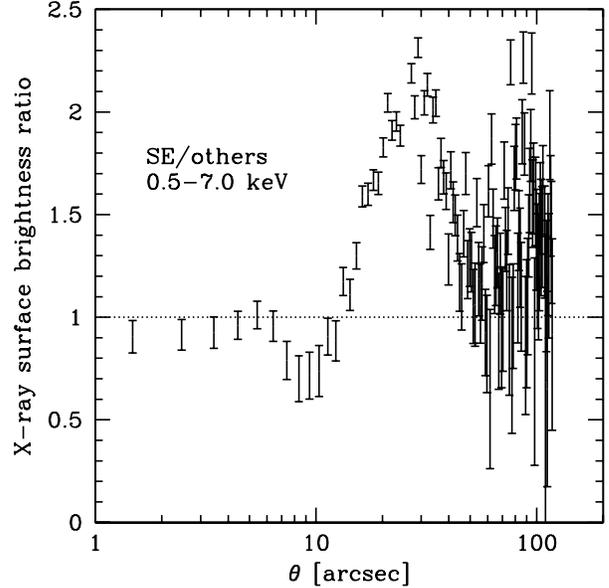}
  \end{center}
  \caption{Ratio of the 0.5--7.0 keV surface brightness in the SE
quadrant and the other directions.  The dotted line corresponds to an
isotropic profile.}  \label{fig:xcount}
\end{figure}

We use essentially the same Chandra ACIS-S3 data in the 0.5--7.0 keV
band as Allen et al. (2002). The surface brightness contours are
overlaid in figure \ref{fig:szmaps}. The imaging data have been
flat-fielded by correcting the effective area of the telescope and
detector with the exposure map.  The background subtraction was
performed using an on-chip background region, located at $\sim 4'-5'$
from the cluster center.  We further plot in figure \ref{fig:xcount} the
ratio of the counts in the SE quadrant and the cluster average
component. The bin size is $0''.984$, corresponding to two detector
pixels. The excess emission has a prominent peak at $\theta \sim 25''$
with very high significance, which coincides with the location of the
excess SZE.  \footnote{The ratios of the counts at the SE peak position
to those of the background are 110 and 16829 for ROSAT HRI (0.2--2 keV)
and Chandra ACIS-S3 (0.5--2 keV), respectively. The significant
improvement in the S/N in Chandra observations turns out to be crucial
for the unambiguous detection of the SE excess, which was not prominent
in the ROSAT HRI image of Schindler et al. (1997).}

\section{Combined Analysis on the Cluster Average Component}
\label{sec:analy}

We first focus on the global properties of RX~J1347--1145, excluding the
SE quadrant. The projected SZE and X-ray data for the cluster average
component both appear to be smooth.  We thus model this component as a
spheroid with an axis ratio of $1 : 1: \eta$, where $\eta$ is an
elongation factor in the line-of-sight direction.

\subsection{Modeling the X-Ray Surface Brightness}
\label{sec:xanaly}

In modeling the X-ray imaging data, we compute the volume emissivity of
a fully ionized plasma with SPEX ver 1.10 (Kaastra et
al. 1996). \footnote{Information available at
http://www.sron.nl/divisions/hea/spex/.} We fix the absorbing hydrogen
column density at the nominal Galactic value, $N_{\rm H}= 4.85 \times
10^{20}$ cm$^{-2}$ (Dickey, Lockman 1990), metallicity at its global
mean for the cluster average component, $Z=0.41$ Z$_\odot$ (Allen et
al. 2002), and the ratio of electron and hydrogen numbers at $n_{\rm
e}/n_{\rm H} = 1.20$ (Feldman 1992).  The model X-ray surface brightness
is then given through a line-of-sight integral of the emissivity, using
electron temperature and density profiles described below.

We assume that the electron temperature at position $(\theta,\phi)$ in
the cluster is given by an axi-symmetric model along the line-of-sight: 
\begin{equation}
T_{\rm e}(\theta,\phi) = T_{\rm e}^{\rm sph} \left[
\sqrt{\theta^2+(\phi/\eta)^2} \right],
\label{eq:tmodel}
\end{equation}
where $\theta$ is an angular radius on the sky, $\phi$ is the
corresponding size in the line-of-sight direction, and $T_{\rm e}^{\rm
sph}$ is the temperature deprojected assuming $\eta=1$.  Unless
otherwise stated, we adopt for $T_{\rm e}^{\rm sph}$ the temperature
profile data from the spectral deprojection analysis presented in figure
5 of Allen et al. (2002). They are also plotted for reference in figure
\ref{fig:tprof} and given in eight radial bins at 0--3.5, 3.5--7.4,
7.4--14.2, 14.2--21.5, 21.5--36.0, 36.0--50.7, 50.7--72.2, and
72.2--108.2 arcsec.

The electron density profile is correspondingly approximated by a
beta-model: 
\begin{equation}
n_{\rm e}(\theta,\phi) = n_{\rm e0}
\left[ 1 + \frac{\theta^2 + (\phi/\eta)^2}{\theta_{\rm c}^2} 
\right] ^{-3\beta/2}, 
\label{eq:betamodel}
\end{equation}
where the central density $n_{\rm e0}$, the angular core radius $\theta_{\rm
c}$, the slope parameter $\beta$, and $\eta$ are free parameters.  The
outer boundary of the cluster is taken at $\sqrt{\theta^2 +
(\phi/\eta)^2} = \theta_{\rm vir}$, where $\theta_{\rm vir}=350''$ is
the angular size corresponding to the virial radius, 2.0 Mpc, of this
cluster (Allen et al. 2002). As the X-ray temperature profile data are
available only within an angular radius of $108''.2$, we use the
temperature in the outermost bin for the regions beyond that.  The
results of our analysis are insensitive to a specific choice of the
boundary, because the X-ray emission declines rapidly at the envelope
(and the signal relative to the map edge is used for the SZE).

We fit the X-ray surface brightness data between radii $4''.9$ and
$108''.2$ with a bin size of $0''.984$, excluding the SE quadrant.  The
model surface brightness from equations (\ref{eq:tmodel}) and
(\ref{eq:betamodel}) is averaged over each radial bin, neglecting the
convolution with the detector point spread function. We then obtain
\begin{eqnarray}
\label{eq:ne0}
n_{\rm e0} ~\eta^{1/2} &=& 0.128^{+0.014}_{-0.013} ~~\mbox{cm}^{-3}, 
\\
\theta_{\rm c} &=& 7''.37^{+0.71}_{-0.62},  
\label{eq:thetac} \\
\beta &=& 0.583^{+0.012}_{-0.012},
\label{eq:beta}
\end{eqnarray}
with $\chi^2=118$ for 102 degrees of freedom (d.o.f.). The data at
$\theta < 4''.9$ show an excess over a simple beta-model profile; a fit
including them yields $\chi^2/{\rm d.o.f.} = 250/107$. This departure is
irrelevant to modeling the SZE data at $\theta >10''$ in the present
analysis.  We attempt to break the degeneracy between $n_{\rm e0}$ and
$\eta$ using the SZE data in subsection \ref{sec:pec}.

In determining the errors quoted in the above equations, we took into
account the errors in the adopted temperature profile and metallicity as
follows. First, the mean values of the beta-model parameters were
determined using the mean values of the temperature in each radial bin
and metallicity. Second, temperatures were varied randomly within their
1-$\sigma$ errors in each radial bin. The metallicity was also varied
within its 1-$\sigma$ error, $\pm 0.17$~Z$_\odot$, to which we assigned
the variance of the radially binned metallicities (table 4 of Allen et
al. 2002), assuming that they are distributed randomly around the global
mean $Z=0.41$~Z$_\odot$.  For a given set of such temperatures and
metallicity, we drew a confidence region ellipsoid in the ($n_{\rm e0}
\eta^{1/2}$, $\theta_{\rm c}$, $\beta$) space. Third, the same procedure
was repeated for different sets of temperatures and metallicity. The
errors of the beta-model parameters were determined by the ``envelope''
of all the confidence-region ellipsoids.  Since correlation of
temperatures between different radial bins or with metallicity was
neglected, the derived errors should be regarded as upper limits on
their true values.

\subsection{Modeling the SZE Surface Brightness}
\label{sec:szanaly}

We adopt the same electron temperature and density profiles as equations
(\ref{eq:tmodel}) and (\ref{eq:betamodel}) to model the SZE.  Also
incorporating the line-of-sight peculiar velocity, $V_{\rm p}$, of a
cluster, the SZE intensity at angular radius $\theta$ and frequency
$\nu$ is computed as
\begin{eqnarray}
&& I_\nu(\theta) = 2 i_0 ~d_{\rm A}(z)
\int_{0}^{\phi_{\rm max}}
\left\{p(x) \left[\frac{kT_{\rm e}(\theta,\phi)}{m_{\rm e} c^2} \right] 
 \sigma_{\rm T} n_{\rm e}(\theta,\phi) \right. \nonumber \\
&& ~+~\left.q(x) \left( \frac{V_{\rm p}}{c} \right) 
\sigma_{\rm T} n_{\rm e}(\theta,\phi) ~
+~ C_{\rm R}(x,  n_{\rm e}, T_{\rm e}, V_{\rm p} ) \right\} ~ d\phi, 
\label{eq:sz}
\end{eqnarray}
where 
\begin{eqnarray}
\label{eq:px}
 p(x) &=& [x \coth(x/2) - 4] ~ q(x), \\
q(x) &=& x^4 e^x/(e^x -1)^2,
\label{eq:qx}
\end{eqnarray}
and $x = h_{\rm P} \nu /(k T_{\rm CMB})$, $i_0 = 2 (k T_{\rm CMB})^3
/(h_{\rm P}c)^2$, $d_{\rm A}(z)$ is the angular diameter distance to the
cluster redshift $z$, $T_{\rm CMB}$ is the CMB temperature, $h_{\rm P}$
is the Planck constant, $m_{\rm e}$ is the electron mass, $\sigma_{\rm
T}$ is the Thomson cross section, $c$ is the speed of light, and
$\phi_{\rm max} = \eta \sqrt{\theta_{\rm vir}^2 - \theta^2}$.  The first
term in the integral of equation (\ref{eq:sz}) corresponds to the
thermal SZE, the second term to the kinematic SZE, and the last term to
their relativistic corrections.  We adopt the correction factor $C_{\rm
R}$ given in Nozawa et al. (1998) and Itoh et al. (2003), applicable to
electron temperatures up to $\sim 50$~keV.  The peculiar velocity $V_{\rm
p}$ is defined to be positive if the cluster is moving toward us.

To take into account the finite spatial resolution of our detectors, the
SZE intensity in equation (\ref{eq:sz}) is convolved with a Gaussian
beam; 
\begin{equation}
I_\nu^{\rm conv}(\theta) = \frac{1}{2\pi \sigma_{\rm beam}^2}
\int  I_\nu(\boldsymbol{\theta}')  
\exp\left[-\frac{(\boldsymbol{\theta}'-\boldsymbol{\theta})^2}{2 
\sigma_{\rm beam}^2} \right] 
d^2\theta',  
\label{eq:conv}
\end{equation}
where $\sigma_{\rm beam}= {\rm FWHM}/\sqrt{8\ln2}$ specifies the
beam-size. The convolved intensity, $I_\nu^{\rm conv}(\theta)$, is then
averaged over a given radial bin to calculate the SZE surface brightness
profile. Geometrical shapes of the observed field (squares at 150~GHz
and 21~GHz) are taken into account at the map edge.  The zero level of
the model profile is defined in the same way as the observed data.

In figures \ref{fig:prof} and \ref{fig:prof_21}, we have plotted the SZE
surface brightness profiles expected from Chandra X-ray imaging and
spectroscopy in the case of $\eta=1$ and $V_{\rm p}=0$.  The expected
SZE signals for the cluster average component are as a whole consistent
with the observed data in all three bands.

\subsection{Elongation Factor and Peculiar Velocity}
\label{sec:pec}

Combining the X-ray and SZE data, one can directly probe the
line-of-sight elongation of a cluster. Multi-band data of the SZE also
give a measure of the cluster bulk motion via the difference between the
spectral shapes of the thermal and kinematic SZE [eqs (\ref{eq:px}) and
(\ref{eq:qx})].

\begin{figure}
  \begin{center}
    \FigureFile(82mm,82mm){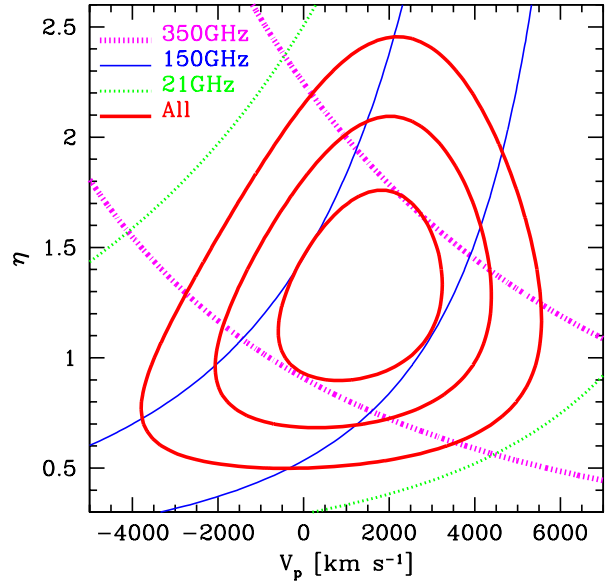}
  \end{center}
  \caption{Constraints on the elongation factor $\eta$ and peculiar
velocity $V_{\rm p}$ for the cluster average component excluding the SE
quadrant.  The thick solid contours indicate the 68.3, 95.4, and 99.7\%
($\Delta \chi^2 = 2.3$, 6.2, and 11.8 for two parameters of interest)
confidence regions from a joint fit to the data in the three bands. The
other contours show the 68.3\% confidence region from each of the
350~GHz (thick dotted), 150~GHz (thin solid), and 21~GHz (thin dotted)
data separately.}  \label{fig:vpeta}
\end{figure}

Figure \ref{fig:vpeta} shows the constraints on the elongation factor
$\eta$ and a peculiar velocity $V_{\rm p}$ from the SZE data at 21, 150,
and 350~GHz.  The beta-model parameters are held fixed at their best-fit
values given in equations (\ref{eq:ne0}), (\ref{eq:thetac}), and
(\ref{eq:beta}). The confidence contours derived from the SZE decrement
data (21 and 150~GHz) are orthogonal to those from the increment data
(350~GHz). The combination of both is therefore essential in deriving
stringent constraints on the parameters. The best-fit values from the
joint fit to the data in three bands are
\begin{eqnarray}
\label{eq:eta}
\eta &=& 1.30^{+0.29}_{-0.26}, \\
V_{\rm p} &=& 1420^{+1170}_{-1270} \mbox{~~km s$^{-1}$},
\label{eq:vp}
\end{eqnarray}
where $\chi^2/{\rm d.o.f.} = 7.3/11$ and the quoted errors correspond to
$\Delta \chi^2 = 1.0$ (68.3\% confidence level for a single parameter
of interest). Given rather large errors, there is only weak indication
for elongation along the line-of-sight or positive bulk velocity of the
cluster. The inferred central $y$-parameter is $y_0 = 9.0 \times 10^{-4}
\eta^{1/2} \simeq 1.0 \times 10^{-3}$. Note that this is an extrapolated
value from the data at $\theta >10''$.

Alternatively, one may assume that the cluster is spherically symmetric
and interpret $\eta$ as being a correction factor for the distance
scale; the angular diameter distance to the cluster, $d_{\rm A}(z)$, is
longer, or equivalently, the Hubble parameter $h$ is smaller, by a
factor of $\eta$ than adopted here ($h=0.71/\eta$).

\begin{figure}
  \begin{center}
    \FigureFile(82mm,82mm){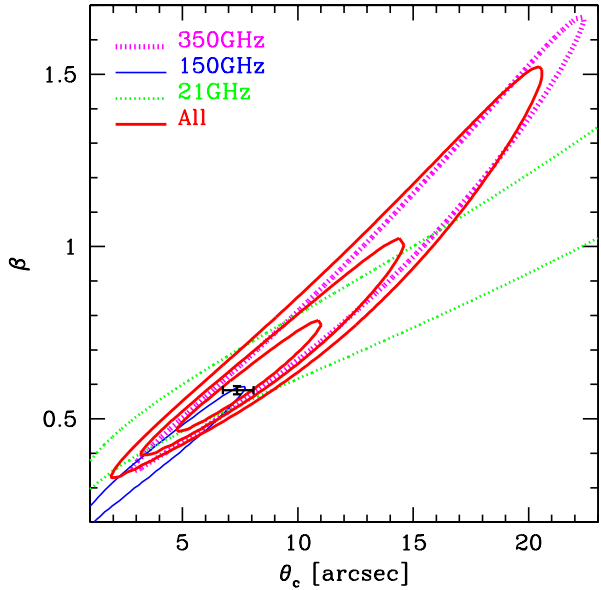}
  \end{center}
\caption{Constraints on the beta-model parameters $\theta_{\rm c}$ and
$\beta$ for the cluster average component excluding the SE quadrant.
The meanings of the contours are the same as in figure \ref{fig:vpeta}.
The error bars are the limits from the X-ray surface brightness data by
Chandra given in equations (\ref{eq:thetac}) and (\ref{eq:beta}).}
\label{fig:rcbeta}
\end{figure}

In order to check the consistency of the shape of the SZE and X-ray
surface brightness profiles more quantitatively, we plot in figure
\ref{fig:rcbeta} the limits on $\theta_{\rm c}$ and $\beta$ from fitting
our SZE data at 21, 150 and 350~GHz. The other parameters $n_{\rm e0}$,
$\eta$ and $V_{\rm p}$ are fixed at their mean values [eqs
(\ref{eq:ne0}), (\ref{eq:eta}), and (\ref{eq:vp})].  The 21~GHz data
yield the weakest constraints on $\theta_{\rm c}$ because only the data
at $ \theta > 60''$ are used.  The best-fit values from a joint fit to
the data in the three bands are
\begin{eqnarray}
\label{eq:thetac2}
\theta_{\rm c} &=& 7''.43^{+2.12}_{-1.96}, \\
\beta &=& 0.588^{+0.114}_{-0.090}, 
\label{eq:beta2}
\end{eqnarray}
where $\chi^2/{\rm d.o.f.} = 7.2/11$ and the quoted errors correspond to
$\Delta \chi^2 = 1.0$.  The quantitative agreement with the X-ray
results further ensures that the electron density profile of the cluster
excluding the SE quadrant and the central $\sim 10''$ is well
approximated by the beta-model.

If we adopt a simplified isothermal model with the emission-weighted
temperature $k T_{\rm e}=12.0^{+0.62}_{-0.59}$~keV, a fit to the X-ray
surface brightness profile between radii $4''.9$ and $108''.2$ yields
$n_{\rm e0}~\eta^{1/2} = 0.135 \pm 0.008 ~\mbox{cm}^{-3}$, $\theta_{\rm
c}=7''.01 \pm 0.27$, and $\beta = 0.580 \pm 0.004$, with $\chi^2/{\rm
d.o.f.} = 115/102$.  The errors include those in the assumed temperature
and metallicity as described in subsection \ref{sec:xanaly}. The above
beta-model parameters are still consistent with equations
(\ref{eq:ne0}), (\ref{eq:thetac}), and (\ref{eq:beta}) within 1-$\sigma$
errors.  The fitted values for $\eta$ and $V_{\rm p}$ in the isothermal
model are $\eta=2.28 ^{+0.52}_{-0.47}$ and $V_{\rm p}=1020
^{+830}_{-920}$ km s$^{-1}$, respectively, with $\chi^2/{\rm d.o.f.} =
8.9/11$.  A comparison with the value in our detailed model
[eq. (\ref{eq:eta})] indicates the importance of the temperature profile
in determining the Hubble constant by the SZE (Inagaki et al. 1995).
The fact that the isothermal approximation in this cluster overestimates
$\eta$ (i.e., underestimates the Hubble constant) is consistent with the
finding of Schmidt, Allen, and Fabian (2003).

\subsection{Temperature Deprojection}

With the current X-ray spectrometers, it is still difficult to determine
reliably the temperature of gas well above 10 keV. It is also likely
that spatially resolved X-ray spectroscopy is unavailable for a sample
of distant clusters targeted for in the future SZE surveys.  We thus
explore a new possibility of constraining the temperature structure of
clusters solely by imaging observations.

By fitting simultaneously the surface brightness profiles of the SZE
($\propto \int n_{\rm e} T_{\rm e} dl$) and X-rays ($\propto \int n_{\rm
e}^2 T_{\rm e}^{1/2} dl$), one can in principle determine the radial
electron density and temperature profiles of a cluster.  We fit the
radial profile data at 150~GHz, 350~GHz and 0.5--7.0 keV of
RX~J1347--1145 with the following set of free parameters: $n_{\rm e}$,
$\theta_c$, $\beta$, and temperatures in five radial bins 10--20,
20--30, 30--40, 40--60, and 60--80 arcsec.  Neglecting for a moment the
knowledge of spatially resolved spectroscopy, the temperature in other
parts of the cluster is fixed at its global mean, $kT_{\rm e} =12.0$
keV.  Given the quality of the present data, we assume $\eta=1$ and
$V_{\rm p}=0$ for simplicity.

\begin{figure}
  \begin{center}
    \FigureFile(82mm,82mm){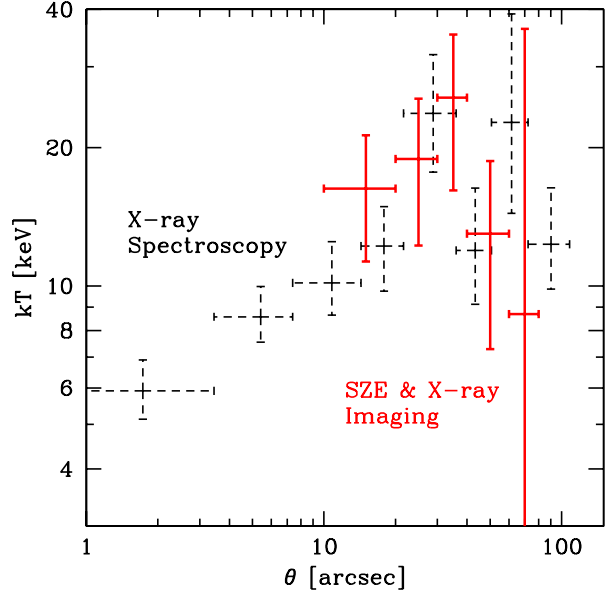}
  \end{center}
  \caption{Deprojected temperatures from a combination of the SZE and
 X-ray imaging data (solid error bars), independent of spatially
 resolved X-ray spectroscopy.  Also plotted for reference are the
 deprojected temperatures from the Chandra X-ray spectroscopy used in
 subsections \ref{sec:xanaly}--\ref{sec:pec} (dashed error bars, Allen
 et al. 2002).}  \label{fig:tprof}
\end{figure}

Figure \ref{fig:tprof} compares the deprojected temperatures in the
current method and the X-ray spectral analysis of Allen et
al. (2002). With the present set of multi-wavelength imaging data, we
are able to constrain the radial temperature profile between $10''$ and
$60''$, which is indeed consistent with the results of the Chandra X-ray
spectroscopy. The fitted values of the other parameters are $n_{\rm
e}=0.130 \pm 0.006$ cm$^{-3}$, $\theta_{\rm c}=7''.34 \pm 0.31$,
$\beta=0.585 \pm 0.005$ with $\chi^2/{\rm d.o.f.}= 121/107$, in good
agreement with equations (\ref{eq:ne0}), (\ref{eq:thetac}), and
(\ref{eq:beta}).

\section{Combined Analysis on the High Temperature Substructure}
\label{sec:se}

A combination of high-resolution SZE and X-ray images further provides a
unique diagnosis of internal structures of galaxy clusters.  We apply
this method to the southeast excess component of RX~J1347--1145 and
determine its mean temperature, gas density, and the extent along the
line-of-sight as follows.  The X-ray and SZE signals are modeled in the
same way as in section \ref{sec:analy}, except that the beta-model
density profile and the temperature profile [eqs (\ref{eq:tmodel}) and
(\ref{eq:betamodel})] are not relevant in this section.
 
First, we assume that the excess component is embedded in the ambient
gas identical to that in the other directions of the cluster, i.e.,
background and foreground components are approximated by the cluster
average component. The signals due to the excess are then given by the
differences in the surface brightness between the SE quadrant and the
cluster average component at 150~GHz, 350~GHz, and 0.5--7.0~keV.  We
take the difference at $\theta = 10''- 28''.6$, where the enhancement is
most evident in both the SZE and X-ray images.  Second, the
spectroscopic temperature in this region $18.0^{+2.7}_{-2.3}$~keV (Allen
et al. 2002) is regarded as an average temperature, weighted by the
detected X-ray count rates, of the excess component and the ambient
gas. For the latter, we assign the temperature of the cluster average
component at the same radius, $12.7 \pm 1.0$~keV.  Third, we assume that
the excess gas is distributed uniformly (like a cylinder) over the
physical length $L_{\rm ex}$ along the line-of-sight with electron
density $n_{\rm ex}$ and temperature $T_{\rm ex}$.  We fiducially fix
the line-of-sight peculiar velocity of the excess gas at the mean value
in equation (\ref{eq:vp}).

From the excess signals at 150~GHz, 350~GHz, and 0.5--7.0~keV, and of the
X-ray emission weighted temperature, we obtain: 
\begin{eqnarray}
\label{eq:te}
kT_{\rm ex} &=& 28.5 \pm 7.3 \mbox{~~keV},\\ 
\label{eq:ne}
n_{\rm ex} &=& (1.49 \pm 0.59) \times 10^{-2} \mbox{~~cm$^{-3}$}, \\
L_{\rm ex} &=& 240 \pm 183 \mbox{~~kpc}, 
\label{eq:l}
\end{eqnarray}
corresponding to the mean $y$-parameter of $4.1 \times 10^{-4}$.
Including the line-of-sight velocity as an additional free parameter
gives only a weak constraint, $V_{\rm ex} = 790 \pm 6880$ km s$^{-1}$,
while the other parameters are essentially unchanged: $kT_{\rm ex} =
28.4 \pm 7.4 \mbox{~keV}$, $n_{\rm ex} = (1.53 \pm 0.73) \times 10^{-2}
\mbox{~cm$^{-3}$}$, and $L_{\rm ex} = 229 \pm 213
\mbox{~kpc}$. Alternatively, by fitting only the surface brightness
differences, without any knowledge of the X-ray emission-weighted
temperature, the results are $kT_{\rm ex} = 33.4 \pm 56.9 \mbox{~keV}$,
$n_{\rm ex} = (1.75 \pm 2.76) \times 10^{-2} \mbox{~cm$^{-3}$}$, and
$L_{\rm ex} = 179 \pm 543 \mbox{~kpc}$. While the errors are still
large, the latter fit indicates a further possibility of constraining
the gas temperature via the relativistic correction of the SZE once the
data attain better precision.  Equations (\ref{eq:ne}) and (\ref{eq:l}),
together with the projected area of the SE substructure on the sky, 560
arcsec$^2$, give the total gas mass of $M_{\rm gas} \sim 2 \times
10^{12}$~M$_{\rm \odot}$ for the excess component.

We have also attempted to constrain the temperature of this excess
component directly with the Chandra X-ray spectra. We first fit the
0.5--7.0 keV spectrum of the cluster average component between radii
$8''.8$ and $28''.6$ with the MEKAL thin-thermal plasma model (Mewe et
al. 1985, 1986; Kaastra 1992; Liedahl et al. 1995) to obtain $kT_{\rm e}
= 12.6^{+1.2}_{-1.0}$~keV, a value consistent with that reported in
Allen et al. (2002). We then fit the spectrum of the SE quadrant of the
same radii with a two-component MEKAL model, incorporating both the
cluster average with 12.6~keV (fixed) and the excess. The amplitude of
the former component was reduced by a geometrical factor of 3. Though
weak, the derived limit on the temperature of the latter component,
$kT_{\rm ex} > 21.5$ keV at the 90\% confidence level, is fully
consistent with equation (\ref{eq:te}).

\section{Discussion}

The presence of very high temperature gas in excess of 20~keV inferred
in section \ref{sec:se} supports strongly an idea that the cluster has
undergone a recent subcluster merger (Cohen, Kneib 2002; Allen et
al. 2002). We explore the properties of this region based on a simple
one-dimensional shock model as follows.

We assume that the pre-shock gas has the density $n_1$ and the
temperature $k T_1= 12.7$ keV expected for the cluster average
component. Given a steep density gradient and an uncertainty in the
subclump position along the line-of-sight, $n_1$ is regarded as being
yet undetermined.  The post-shock gas is assumed to have $n_2=1.49
\times 10^{-2}$ cm$^{-3}$ and $kT_2 = 28.5$ keV from equations
(\ref{eq:te}) and (\ref{eq:ne}).  Applying the Rankine--Hugoniot
conditions across the shock with an adiabatic index $\gamma=5/3$, we
find that the Mach number of the pre-shock gas is ${\cal M}_1=2.1$, and
the velocity of the pre-shock and post-shock gas relative to the shock
front are $v_1=3900$~km~s$^{-1}$ and $v_2=1600$~km~s$^{-1}$,
respectively.  For a head-on collision of equal mass clumps, the
collision velocity is $\Delta v = 2(v_1-v_2)=4600$~km~s$^{-1}$. The
corresponding density ratio $n_2/n_1=2.4$ implies $n_1= 6.2 \times
10^{-3}$ cm$^{-3}$. Assuming further that the pre-shock gas distribution
is described by equations (\ref{eq:betamodel})--(\ref{eq:beta}) with
$\eta \sim 1$, the distance of the subclump from the cluster center,
including the line-of-sight displacement, is suggested as
$\sqrt{\theta^2 + \phi^2} \sim 40''$ or 230~kpc.

Previous hydrodynamical simulations of Takizawa (1999), for instance,
showed that the gas temperature can rise up to $kT_{\rm e} > 25$ keV as
a result of a head-on collision of $5\times 10^{14}$~M$_\odot$ clusters
with relative velocity $\Delta v >4000$ km s$^{-1}$. The highest
temperature gas is likely to emerge near the bounce-shock with total gas
mass of $\sim 10^{12}$~M$_\odot$ for a duration of $\sim 0.5$~Gyr
(M.~Takizawa, private communication).  Gravitational softening length of
20~kpc and the gas particle mass of $10^{10}$ M$_\odot$ used in the
simulations are sufficient for resolving the substructures comparable to
that found in RX~J1347--1145. Though it is still premature to conclude
the origin of the hot gas in RX~J1347--1145, these results are consistent
with our interpretation of shock heating during a violent merger.

A merger event may also produce high-energy nonthermal particles. While
BeppoSAX observations of RX~J1347--1145 (Ettori et al.  2001) have found
little evidence for their contribution to the entire X-ray emission of
this cluster, nonthermal particles may affect the enhanced SZE and X-ray
signals locally. We examined the Chandra spectrum of the SE excess
position ($8''.8-28''.6$), and found that the effective photon index is
1.4 without subtracting the cluster average component.  This is
comparable to, or even harder than, those expected from the inverse
Compton and nonthermal bremsstrahlung emissions, 1.5 and 2.0,
respectively, in the strong shock limit (Sarazin 1999; Sarazin, Kempner
2000).  Thus, if the nonthermal component is included in the spectral
analysis, the inferred thermal temperature would become higher, unless
the spectrum is fully dominated by the inverse Compton emission
resulting from the strong shock. A harder nonthermal electron spectrum
may be possible as a result, for instance, of Coulomb losses, but they
will also lead to unacceptably large heating of the thermal gas
(Petrosian 2001; Takizawa 2002).  Contribution of nonthermal electrons
to the SZE signals depends sensitively on the amount and spectrum of
relatively low energy nonthermal electrons (e.g., Ensslin, Kaiser 2000;
Colafrancesco et al.  2003), but is unlikely to dominate over the
thermal SZE (Shimon, Rephaeli 2002). In any case, if the nonthermal
component is separated from the thermal and kinematic SZE in future
observations, it will provide a useful probe of the nature of
high-energy populations in clusters.

In the present analysis, we excluded the regions around already resolved
contaminating sources.  One further possibility of contamination,
particularly in the 350~GHz image, is unresolved emission from
background submillimeter sources (Smail et al. 2002 and references
therein).  If such sources are distributed rather uniformly in the
cluster field, their average contribution has been subtracted in our
analysis, as we only used the signal relative to the map
edge. Shot-noise-like signals, on the other hand, may increase the noise
level of the image, but are unlikely to produce the extended feature
illustrated in figure \ref{fig:quad}. The gravitational lensing effect
tends to enhance the emission close to the critical lines, not toward
the cluster center.  We thus conclude that the contribution of the
background sources is unlikely to change the main results of this paper.

Another possibility of contamination is dust emission inside the
cluster. Edge et al. (1999) report detection of such emission in two out
of seven central galaxies in clusters. Dust may also be present in the
intracluster medium (Dwek et al.  1990; Stickel et al. 2002), although
such diffuse dust is likely to be destroyed via sputtering within the
age of the Universe (Draine, Salpeter 1979). Since dust emission from
the cluster member galaxies or the intracluster medium will peak at
far-infrared wavelengths, it can be constrained directly by infrared
facilities, such as SIRTF and Astro-F. We will investigate these points
in detail elsewhere (K.~Yamada, T.~Kitayama, in preparation). It will
also be interesting to see if the secondary cD really contains dust, as
implied by our SCUBA map.

\section{Conclusions}

We have performed combined analyses of the highest resolution SZE and
X-ray images currently available for RX~J1347--1145.  The data at both
150~GHz and 350~GHz show clear signatures of enhanced SZE signals at
$\sim 150$~kpc southeast from the center of this cluster.  We have
separately determined the temperature, density, and the line-of-sight
extent of this excess component, after removing the foreground and
background signals.  The presence of high-temperature gas with $28 \pm
7$~keV strongly indicates that the cluster has recently experienced a
major merger.

Excluding the regions around the southeast enhancement and the radio
sources, the SZE signals at 350, 150, and 21~GHz agree well with those
expected from the electron density and temperature profiles of Chandra
X-ray observations.  Our multi-band SZE data indicate marginal evidence
for the line-of-sight elongation of the gas distribution or non-zero
peculiar velocity.  The central $y$-parameter extrapolated from radii
greater than $60$~kpc is $y_0 \simeq 1.0 \times 10^{-3}$.  We further
present a temperature deprojection technique using the SZE and X-ray
imaging data. The radial temperature profile so obtained between 60 and
350~kpc are all consistent with those based on the X-ray spectral
deprojection analysis of Allen et al. (2002).

Our results demonstrate the power of high-resolution SZE mapping
observations. In particular, they provide a unique probe of the
temperature structure in distant clusters for which spatially resolved
X-ray spectroscopy is unavailable.  With its unlimited sensitivity to
high temperature, the SZE has an advantage over X-rays in detecting the
gas with temperatures well above 10 keV. As demonstrated for
RX~J1347--1145, an angular resolution of $\sim 10''$ is essential in 
resolving irregular morphology of merger-heated gas in distant clusters.
Large detector arrays, such as BOLOCAM on the Large Millimeter Telescope
(Glenn et al. 2003) and SCUBA-2 on the James-Clerk-Maxwell Telescope
(Holland et al. 2003), will indeed be suitable for such observations.
Though performed on a single cluster so far, the methodology presented
in this paper will be applicable to a statistical sample of clusters
obtained in the future SZE surveys.

\bigskip 

We thank Steve Allen for providing deprojected temperature data and for
useful discussions, Naoki Itoh and Satoshi Nozawa for supplying
subroutines of their relativistic correction formulae, and Erik Reese,
Motokazu Takizawa, Ichi Tanaka and the referee for helpful comments and
discussions. NO and KY acknowledge support from Research Fellowships of
the Japan Society for the Promotion of Science for Young Scientists.
This work is supported in part by the Grant-in-Aid by the Ministry of
Education, Culture, Sports, Science and Technology (14740133, TK).

\section*{References}
\small

\re Allen, S. W., Schmidt, R. W., \& Fabian, A. C. 2002, MNRAS, 335, 256

\re Bennett, C. L., et al.  2003, ApJS, 148, 1

\re Birkinshaw, M.\ 1999, Phys. Rep. 310, 97

\re Carlstrom, J. E., Holder, G. P., \& Reese, E. D. 2002, ARA\&A, 40, 643

\re Carlstrom, J. E., Joy, M., \& Grego, L. 1996, ApJ, 456, L75

\re Cohen, J. G., \& Kneib, J. 2002, ApJ, 573, 524

\re Colafrancesco, S., Marchegiani, \& P., Palladino, E. \ 2003, A\&A,
397, 27

\re Cooray, A. R., Grego, L., Holzapfel, W. L., Joy, M., \& Carlstrom,
J. E. \ 1998, AJ, 115, 1388

\re Dickey, J. M., \& Lockman, F. J. \ 1990, ARA\&A, 28, 215

\re Donahue, M., Gaskin, J. A., Patel, S. K., Joy, M., Clowe, D., \& 
Hughes, J. P. \ 2003, astro-ph/0308024

\re Draine, B. T., \& Salpeter, E. E. \ 1979, ApJ, 231, 438

\re Dwek, E., Rephaeli, Y.,  \& Mather, J.C. 1990, ApJ, 350, 104

\re Edge, A. C., Ivison, R. J., Smail, I., Blain, A. W., \& Kneib J. -P.\
1999, MNRAS 306, 599

\re Emerson, D. T., \& Gr\"ave, R. \ 1988, A\&A, 190, 353

\re Ensslin, T. A., \& Kaiser, C. R. \ 2000, A\&A, 360, 417

\re Ettori, S., Allen, S. W., \& Fabian, A. C. \ 2001, MNRAS, 322, 187

\re Fabian, A. C., Sanders, J. S., Allen, S. W., Crawford, C. S., 
Iwasawa, K., Johnstone, R. M., Schmidt, R. W., \& Taylor, G. B. \ 2003,
MNRAS, 344, L43

\re Feldman, U. \ 1992, Phys. Scr., 46, 202

\re Glenn, J., et al. \ 2003, Proc. SPIE, 4855, 
ed. T. G. Phillips \& J. P. Zmuidzinas, 30

\re Holland, W. S., et al. \ 1999, MNRAS, 303, 659

\re Holland, W. S., Duncan, W., Kelly, B. D., Irwin, K. D., Walton,
A. J., Ade, P. A. R., \& Robson, E. I. 2003, in Proc. SPIE, 4855,
ed. T. G. Phillips \& J. P. Zmuidzinas, 1

\re Hughes, J. P., Birkinshaw, M.\ 1998, ApJ, 501, 1

\re Inagaki, Y., Suginohara, T., \& Suto, Y. \ 1995, PASJ, 47, 411

\re Itoh, N., \& Nozawa, S. \ 2003, A\&A, submitted  (astro-ph/0307519) 

\re Jenness, T., \& Lightfoot, J. F. \ 1998, in A.S.P. Conference
Series, 145,  ed. R. Albrecht, R.N. Hook and H.A. Bushouse, 216 

\re Jenness, T., Lightfoot, J. F., \& Holland, W. S. \ 1998, in Proc. SPIE,
3357, ed. T. G. Phillips, 548

\re Jones, M., et al. 1993, Nature, 365, 320

\re Kaastra, J. S. \ 1992, An X-Ray Spectral Code for Optically Thin Plasmas
       (Internal SRON-Leiden Report, updated version 2.0)

\re Kaastra, J. S., Mewe, R., \& Nieuwenhuijzen, H. \ 1996, in UV and X-ray
spectroscopy of astrophysical and laboratory plasmas, ed. K. Yamashita
and T. Watanabe, 411

\re Komatsu, E., Kitayama, T., Suto, Y., Hattori, M., Kawabe, R.,
Matsuo, H., Schindler, S., \& Yoshikawa, K.\ 1999, ApJ, 516, L1

\re Komatsu, E., et al. 2001, PASJ, 53, 57

\re Kuno, N., Matsuo, H., Mizumoto, Y., Lange, A. E., Beeman, J. W.,
\& Haller, E. E. \ 1993, Int. J. Infrared Millimeter Waves, 14, 749

\re Liedahl, D.A., Osterheld, A.L., \& Goldstein, W.H. \ 1995, ApJ, 438,
L115

\re Markevitch, M., et al. \ 2000, ApJ, 541, 542

\re Markevitch, M., Gonzalez, A. H., David, L., Vikhlinin, A., Murray,
S., Forman, W., Jones, C., \& Tucker, W. \ 2002, ApJ, 567, L27

\re Mewe, R., Gronenschild, E. H. B. M., \& van den Oord, G. H. J. \ 1985, 
A\&AS, 62, 197 

\re Mewe, R., Lemen, J. R., \& van den Oord, G.H.J. 1986, A\&AS, 65, 511

\re Nozawa, S., Itoh, N., \& Kohyama, Y.\ 1998, ApJ, 507, 530

\re Peterson, J. R., Kahn, S. M., Paerels, F. B. S., Kaastra, J. S., Tamura,
T., Bleeker, J. A. M., Ferrigno, C., \& Jernigan, J. G. \ 2003, ApJ, 590, 207

\re Petrosian, V. 2001, ApJ, 557, 560

\re Pointecouteau, E., Giard, M., Benoit, A., D\'esert, F. X., Aghanim,
N., Coron, N., Lamarre, J.M. \& Delabrouille, J. 1999, ApJ, 519, L115

\re Pointecouteau, E., Giard, M., Benoit, A., D\'esert, F. X., 
Bernard J.P., Coron, N., \& Lamarre, J. M. \ 2001, ApJ, 552, 42

\re Pointecouteau, E., Hattori, M., Neumann, D. M., Komatsu, E., Matsuo,
H., Kuno, N., \& B\"ohringer, H. \ 2002, A\&A, 387, 56

\re Reese, E. D., Carlstrom, J. E., Joy, M., Mohr, J. J., 
Grego, L., \& Holzapfel, W. L. 2002, ApJ, 581, 53

\re Rephaeli, Y.\ 1995, ARA\&A, 33, 541

\re Sarazin, C. L.\ 1999, ApJ, 520, 529

\re Sarazin, C. L., \& Kempner, J. C.  2000, ApJ, 533, 73

\re Schindler, S., et al.\ 1995, A\&A 299, L9

\re Schindler, S., Hattori, M., Neumann, D. M., \& B\"ohringer, H.\ 1997, A\&A
317, 646

\re Schmidt, R. W., Allen, S. W., \& Fabian, A. C.  2003, MNRAS submitted 

\re Shimon, M., \& Rephaeli, Y.  2002, ApJ, 575, 12

\re Smail, I., Ivison, R.J., Blain, A. W., \& Kneib, J.-P. 2002, MNRAS,
331, 495

\re Spergel, D. N., et al. 2003, ApJS, 148, 175

\re Stickel, M., Klaas, U., Lemke, D., \&  Mattila, K.  2002, A\&A, 383,
367 

\re Sunyaev, R. A., \& Zel'dovich Ya. B.\ 1970, Ap\&SS, 7, 3

\re Sunyaev, R. A., \& Zel'dovich Ya. B.\ 1972, Comments Astrophys. Space
Phys.,  4, 173

\re Takizawa, M. 1999, ApJ, 520, 514 

\re Takizawa, M. 2002, PASJ, 54, 363 

\re Tsuboi, M., Miyazaki, A., Kasuga, T., Matsuo, H., \& Kuno, N. 1998, 
PASJ, 50, 169 

\re Tsuboi M., Miyazaki A., Kasuga T., Kuno N., Sakamoto A., \& Matsuo
H. 2002, PASJ submitted

\re Yoshikawa, K., \& Suto, Y.  1999, ApJ, 513, 549

\re Zaroubi, S., Squires, G., Hoffman, Y., \& Silk, J.  1998, ApJ, 500, L87

\re Zemcov, M., Halpen, M., Borys, C., Chapman, S., Holland, W.,
Pierpaoli, E., \& Scott, D. 2003, MNRAS, 346, 1179

\end{document}